\documentclass[letterpaper, 10 pt, conference]{ieeeconf}  % Comment this line out if you need a4paper

\IEEEoverridecommandlockouts                              % This command is only needed if 
\usepackage{graphicx} % for pdf, bitmapped graphics files
\usepackage{subcaption}
\usepackage[cmex10]{amsmath} % assumes amsmath package installed
\usepackage{amssymb}  % assumes amsmath package installed
\usepackage{color}
\usepackage{bbm}
\usepackage{algorithm,algorithmicx,algpseudocode}
\usepackage{url}

\usepackage{float}

\usepackage{epstopdf}

\usepackage{tikz}
\usetikzlibrary{automata,positioning}

\makeatletter

\newcommand\floatc@myruled[2]{{\@fs@cfont #1} #2\par}
\newcommand\fs@myruled{\def\@fs@cfont{\bfseries}\let\@fs@capt\floatc@myruled
\def\@fs@pre{\hrule height.8pt depth0pt \kern2pt}%
\def\@fs@post{\kern2pt\vspace{-10pt}\hrule\relax}%
\def\@fs@mid{\kern2pt\hrule\kern2pt}%
\let\@fs@iftopcapt\iftrue}
  
\floatstyle{myruled}
\newfloat{Problem}{tbp}{prob}

\usepackage[official]{eurosym}

\usepackage{mathtools}

\title{\LARGE \bf
Stackable vs Autonomous Cars for Shared Mobility Systems: a Preliminary Performance Evaluation*
}

\author{Chiara Boldrini$^{1}$, Raffaele Bruno$^{1}$ %\\
\thanks{*This work was partially funded by the ESPRIT project. This project has received funding from the \emph{European Union's Horizon 2020 research and innovation programme} under grant agreement No 653395. This work was also partially funded by the REPLICATE project. This project has received funding from the \emph{European Union's Horizon 2020 research and innovation programme} under grant agreement No 691735.}% <-this % stops a space
\thanks{$^{1}$Chiara Boldrini and Raffaele Bruno are with the IIT institute of the Italian National Research Council, Via G. Moruzzi 1, 56124 Pisa, Italy.
        Email: {\tt\small first.last@iit.cnr.it}}
}

\begin{document}

\maketitle
\thispagestyle{empty}
\pagestyle{empty}

%%%%%%%%%%%%%%%%%%%%%%%%%%%%%%%%%%%%%%%%%%%%%%%%%%%%%%%%%%%%%%%%%%%%%%%%%%%%%%%%
\begin{abstract}
Car sharing is one of the key elements of a Mobility-on-Demand system, but it still suffers from several shortcomings, the most significant of which is the fleet unbalance during the day. What is typically observed in car sharing systems, in fact, is a vehicle shortage in so-called hot spots (i.e., areas with high demand) and vehicle accumulation in cold spots, due to the patterns in people flows during the day. In this work, we overview the main approaches to vehicle redistribution based on the type of vehicles the car sharing fleet is composed of, and we evaluate their performance using a realistic car sharing demand derived for a suburban area around Lyon, France. The main result of this paper is that stackable vehicles can achieve a relocation performance close to that of autonomous vehicles, significantly improving over the no-relocation approach and over traditional relocation with standard cars.
\end{abstract}

% Car sharing enables the shift from privately owned cars to shared vehicles, which, as a result, is expected to reduce the overall number of circulating cars, to increase usage of public transport and of non-motorised modes in general. Despite this appealing promises, car sharing

%
%%%%%%%%%%%%%%%%%%%%%%%%%%%%%%%%%%%%%%%%%%%%%%%%%%%%%%%%%%%%%%%%%%%%%%%%%%%%%%%%
%
\section{Introduction}
\label{sec:intro}
\noindent
Mobility-on-Demand systems are a new intermodal mobility concept, also declined as Mobility-as-a-service. The key idea is that people will not own their private car (and be stuck with it) anymore. Instead they will have a fully-fledged choice of transportation modes, seamlessly integrated with each other and with the \emph{smart} city thanks to ICT, automation, and big data. When it comes to motorised modes for personal mobility, car sharing is a staple of MoD systems. Car sharing's positive effects have already been measured: car sharing members use cars less, rely more on public transport, and in some cases they even shed their private car (or refrain from buying a second one for their family)~\cite{martin2016impacts}. Car sharing can also act as a last-kilometre solution for connecting people with public transport hubs, hence becoming a feeder to traditional public transit~\cite{shaheen2001commuter}.

Initial car sharing policies offered limited flexibility to the customers, forcing them to bring the shared cars back to the starting point of their journey. This is the case of two-way car sharing. The recent growth in car sharing popularity and usage, though, has been the result of the availability of car sharing solutions more suitable to customers' needs in terms of flexibility. One-way car sharing, in fact, allows customers to pick up and drop off vehicles at any of the car sharing stations deployed in the city (for station-based systems like Autolib in Paris) or at any location with operator's service area (for free-floating systems like Car2go). Unfortunately, this freedom is often paid in terms of vehicle availability. In fact, cars will follow the natural flows of people in a city, hence accumulating in commercial/business areas in the morning and in residential areas at night~\cite{boldrini16characterising}. As a result, the availability of cars can become extremely unbalanced during the day, and certain areas of high demand (\emph{hot spots}) may end up being underserved while areas with low demand (\emph{cold spots}) may have several cars idly parked and that nobody wants to pick up.

Previous research has proposed several approaches to solve the vehicle unbalance problem, including: user-based relocation, i.e., price incentives for the users to relocate the vehicles themselves~\cite{trr12_relocation}; operator-based relocation, i.e., workforce that moves vehicles from where they are not needed to where there is a significant demand~\cite{Kek2009,Boyaci2015,Nourinejad2015,Weikl2013}; and optimal planning of station deployment to achieve better service accessibility and a more favourable distribution of vehicles~\cite{itcs16_biondi}. It is important to point out that \emph{the relocation process is intrinsically inefficient}: as one driver per car is needed, for relocating several cars a large workforce or many willing customers are necessary. Autonomous cars can be a breakthrough from the car sharing perspective: the  relocation workforce can be shed, as vehicles will move proactively and autonomously where they are needed.

\begin{figure}[t]
\begin{center}
\includegraphics[scale=0.36]{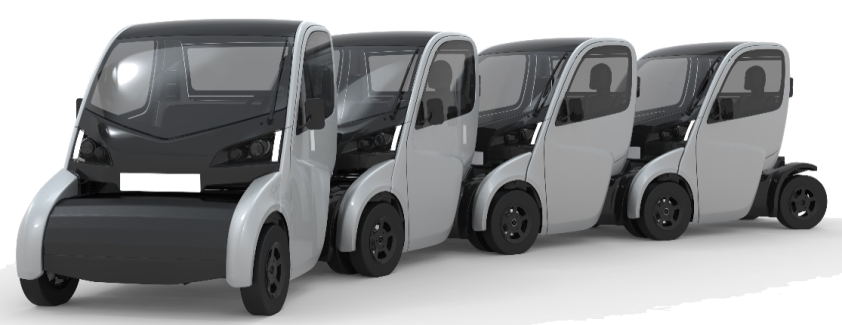}
\caption{The ESPRIT train of vehicles}
\label{fig:esprit}
\end{center}\vspace{-15pt}
\end{figure}

Autonomous vehicles have been the subject of extensive research in the recent years. The first pilot programs have highlighted the great potential of the technology, while also causing some headaches~\cite{nyt-guide-challenges}. While the fact that self-driving cars are going to replace traditional vehicles has never been in question, when exactly they are going to be in people's hands is not clear, with predictions ranging from a few years to a few decades. In the meantime, alternative, intermediate solutions should be sought after to address the problems of currently available transportation systems. As far as car sharing is concerned, new vehicle concepts with \emph{stackable capabilities} have been recently released or are under development, which can be stacked into a train (through a mechanical and electric coupling) and/or folded together. Then, the train can be driven either by a car sharing worker or by a customer. An illustration of this type of vehicle prototyped within the ESPRIT project~\cite{esprit} is provided in Fig.~\ref{fig:esprit}. 

Such stackable cars come with the promises of significantly improving the system manageability of future car sharing services and of being market-ready in just a few years. As such, they could be an effective intermediate solutions before autonomous cars become a reality. It is therefore important to understand how they rank  with respect to future and currently available vehicles. For this reason, in this paper we identify and compare different strategies for vehicle redistribution based on the type of vehicle used. Relying on a realistic car sharing demand, we study how well each of them is faring and we highlight their limitations. 
First, we apply a theoretical upper bounds about relocation with self-driving cars developed in~\cite{ijrr12_fluid} to determine the minimum fleet size that would allow to perfectly rebalanced the considered demand. Second, we developed simple heuristics to quantify the fleet unbalance over time and to define periodic relocation tasks that could help in rebalancing the system. Then we discuss how to perform these relocation tasks with and without operators, using conventional cars or stackable cars. The results show that stackable vehicles can significantly improve the quality of service provided by the car sharing operator with respect traditional cars, and hence should be considered as a viable, intermediate option on the road to self-driving vehicles.

%%%%%%%%%%%%%%%%%%%%%%%%%%%%%%%%%%%%%%%%%%%%%%%%%%%%%%%%%%%%%%%%%%%%%%%%%%%%%%%%
%
%\section{Related work}
%\label{sec:rel_work}
%\noindent
%%
%TBD
%%%%%%%%%%%%%%%%%%%%%%%%%%%%%%%%%%%%%%%%%%%%%%%%%%%%%%%%%%%%%%%%%%%%%%%%%%%%%%%%
%
\section{The car sharing demand}
\label{sec:demand}
\noindent
In this work we consider the car sharing demand obtained from a population synthesiser and demand model developed within the ESPRIT project for the suburban area of Lyon, France \cite{laarabi2017performance-one-way}. This demand corresponds to the potential demand, i.e., the demand obtained from the mobility needs of  the people in the study area and their propensity towards using car sharing. The population synthesiser takes into account a set of 37,402 facilities (i.e., sources of mobility requests, such as households, commercial buildings, PT stops, etc.) scattered in the area, while the demand model determines the pickup and drop-off locations for each car sharing trip request. In this work, we focus on a station-based system, thus we need to deploy stations in the study area in order to associate the car sharing requests to the closest station. The definition of an optimal deployment algorithm is out of the scope of the paper, and for this reason we consider a straightforward deployment whereby we divide the study area using a grid with cells of side length $1$km and we place a station in the centroid of the cell if there is at least one facility in that cell. This implies that an hypothetical customer will find at least one station at a $\sim500$m distance. After this procedure is completed, we end up with 70 active station in the study area. %The map in Figure~\ref{fig:demand_map} shows the study area with all the requests (red dots) issued during a day and the deployed stations. 

\begin{figure}[t]
\begin{center}
\includegraphics[scale=0.4]{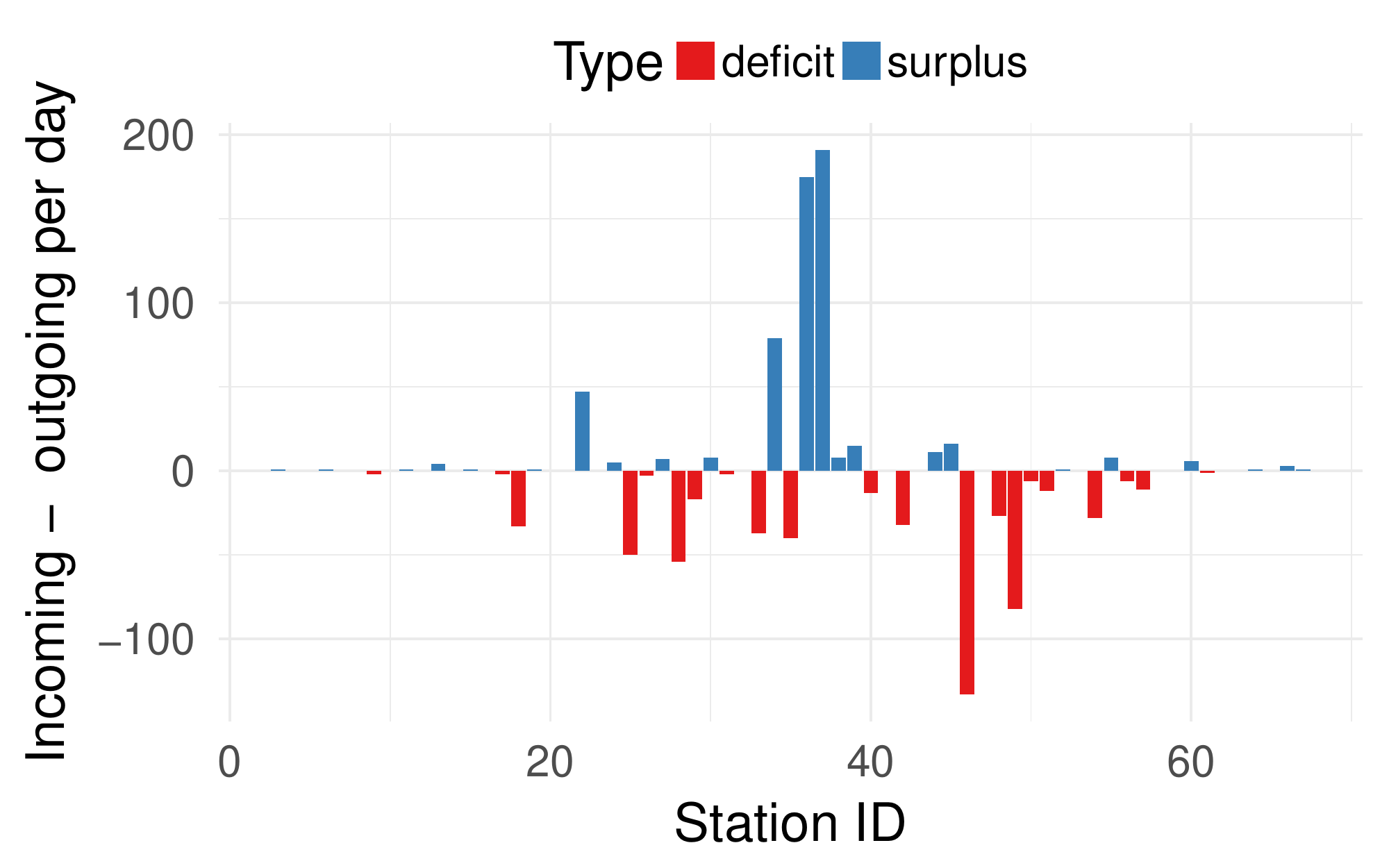}
\caption{Daily unbalance at stations.}
\label{fig:deficit_surplus_per_day}
\end{center}\vspace{-20pt}
\end{figure}

%Figure~\ref{fig:ts_requests} shows the evolution in time of the requests, highlighting the characteristics early morning and late afternoon peaks related to commuting patterns. 
The demand generated by the model will provide the basis for our evaluation in Section~\ref{sec:evaluation}. As in all car sharing systems, stations are used very differently by the customers depending on the time of the day and where they are located~\cite{boldrini16characterising}. This is confirmed in Figure~\ref{fig:deficit_surplus_per_day}, where we focus on individual stations and we compute the difference between their inflow and outflow, which correspond, respectively, to the number of incoming vs outgoing vehicles per day. We observe that there are station that see more drop-offs than pickups, and vice versa. 

%\begin{figure}[h]
%\begin{center}
%\includegraphics[scale=0.5]{figures/demand_with_stations_map.pdf}\vspace{-10pt}
%\caption{Map of the demand, the zoning system, and the stations in the study area.}
%\label{fig:demand_map}
%\end{center}
%\end{figure}

%\begin{figure}[h]
%\begin{center}
%\includegraphics[scale=0.4]{figures/facilities_stations_map.pdf}
%\caption{Map of the facilities, zoning system, and stations in the study area.}
%\label{fig:demand_map}
%\end{center}
%\end{figure}

%\begin{figure}[h]
%\begin{center}
%\includegraphics[scale=0.4]{figures/ts_requests.pdf}
%\caption{Time series of pickup/drop-off requests over one typical working day.}
%\label{fig:ts_requests}
%\end{center}\vspace{-20pt}
%\end{figure}

%%%%%%%%%%%%%%%%%%%%%%%%%%%%%%%%%%%%%%%%%%%%%%%%%%%%%%%%%%%%%%%%%%%%%%%%%%%%%%%%
%
\section{Relocation: current solutions}
\label{sec:relocation_current}
\subsection{No relocation}\label{sec:no_relocation}
\noindent
Relocation can be very costly for car sharing operators, and this is the reason why most of them do not implement redistribution policies in their systems. For example, Car2go has recently announced that only one year ago they have designed their relocation strategy, and that since then they have successively introduced it in all their locations~\cite{car2go_relocation}. Given the practical relevance, the no-relocation case will serve as the baseline for our discussion. In particular, it is interesting to consider how many cars one would need to satisfy the demand in Section~\ref{sec:demand} without ever dropping one request. This number can be computed by simply deriving the number of cars needed in the system at midnight such that the availability at the station is never negative during the following day. Using this approach, we found that we would need 1542 shared vehicles in order to accept all pickup requests in the Lyon study area. 

%%%%%%%%%%%%%%%%%%%%%%%%%%%%%%%%%%%%%%%%%%%%%%%%%%%%%%%%%%%%%%%%%%%%%%%%%%%%%%%%
%
\subsection{Relocation with standard cars}\label{sec:relocation_standard_cars}
\noindent
When the shared fleet is composed of standard cars, relocation can only be performed moving one vehicle at a time. The car sharing workers (the \emph{relocators}) can either work in pairs or alone. When in pairs, one drives the other one (using a service car) to the location where there is shared car to be relocated, then drives again to the relocation endpoint to pick up the relocator. An alternative solution is for the relocator to use a folding bicycle to reach the car to be relocated, then to store the folded bike in the trunk, to drive the car where it is need, then cycle again to reach the car of the following relocation task~\cite{Bruglieri2014}. This process is clearly not very efficient. In the first case, two workers are needed to relocate one car. In the second case, the relocation delays are long since it takes some time for the relocators to reach the next car by bike. 

% When it comes to relocation, the research problem can be separated into two issues: i) prediction of the demand (i.e., identifying hot and cold spots in the network), and ii) implementation of the relocation decisions. 

%%%%%%%%%%%%%%%%%%%%%%%%%%%%%%%%%%%%%%%%%%%%%%%%%%%%%%%%%%%%%%%%%%%%%%%%%%%%%%%%
%
\subsection{Theoretical upper-bound}\label{sec:relocation_theo_bound}
\noindent
As far as relocation is concerned, you can never do better than with autonomous cars. In fact, assuming that a demand prediction tool is in place, self-driving cars can autonomously dispatch themselves where they are needed. %We can thus safely state that relocation with autonomous cars represents the upper bound on the relocation performance, given a certain demand prediction tool. 
Pavone et al. have derived in~\cite{ijrr12_fluid} a theoretical results about relocation with self-driving cars. Their model is based on a representation of the trips in the car sharing network in terms of flows in a fluid model. Pavone et al. prove that, without vehicle redistribution, the car sharing system cannot reach equilibrium (i.e., customers waiting for cars will accumulate indefinitely at stations). Instead, with a redistribution policy in place and a sufficient number of cars, it is always possible to stabilise the car sharing system. Pavone et al. also define an optimisation problem to derive the optimal relocation flows. Clearly, the fluid model is just an approximation of a real car sharing system, but it is very useful and important in order to establish what, in practice, is a theoretical upper-bound on vehicle rebalancing.

\begin{figure}[h]
\begin{center}
\hspace{-15pt}\includegraphics[scale=0.4]{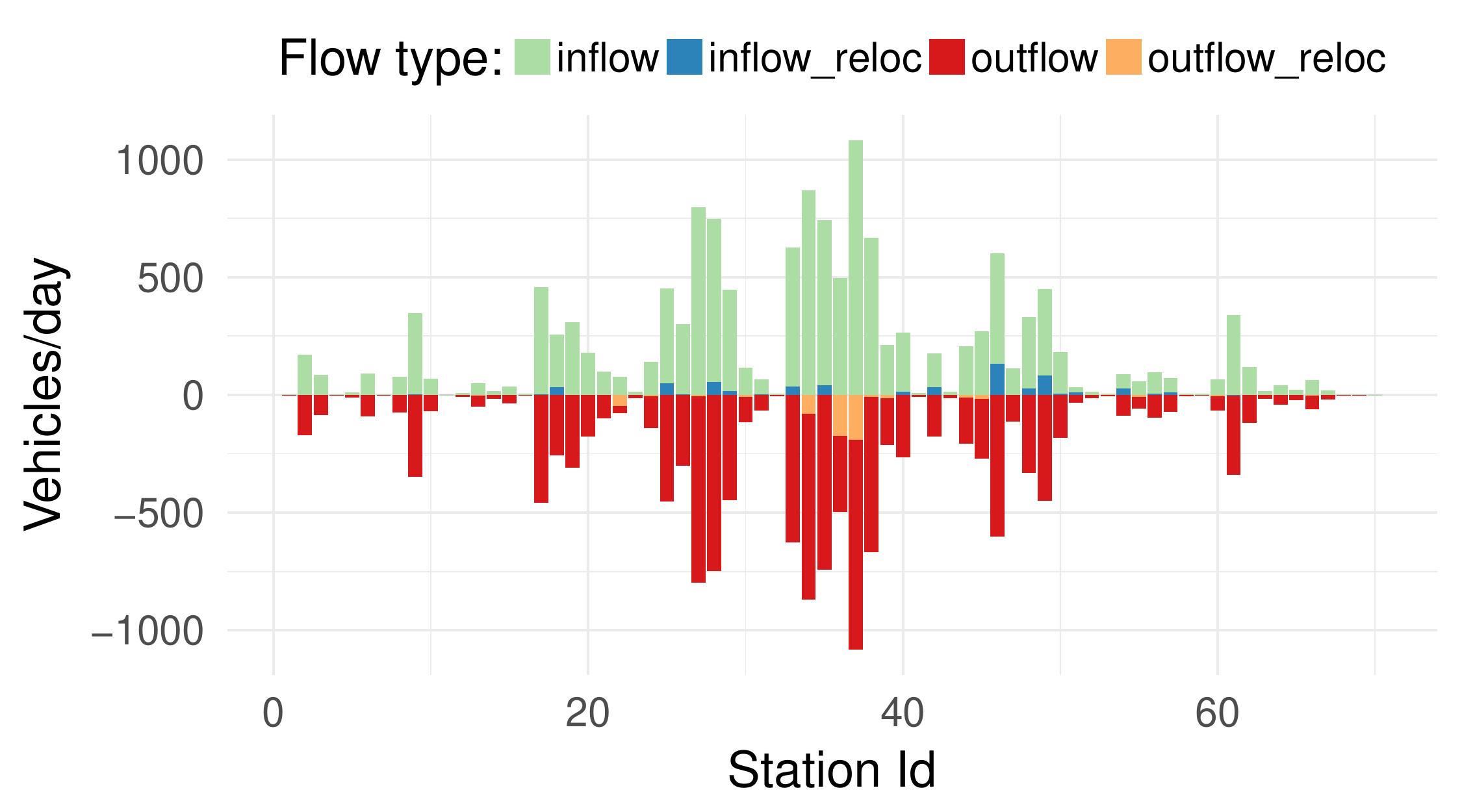}
\caption{Daily unbalance at stations.}
\label{fig:theo_bound}
\end{center}%\vspace{-20pt}
\end{figure}

We apply the model presented in~\cite{ijrr12_fluid} in order to derive the optimal rebalancing flows with autonomous vehicles for the car sharing demand described in Section~\ref{sec:demand}. The results are shown in Figure~\ref{fig:theo_bound}. The blue and orange bars show the inbound/outbound relocation flows from each station. According to the model in~\cite{ijrr12_fluid}, optimal rebalancing is possible using 292 vehicles. Please note that this is 5 times less than what it would be needed without relocation (Section~\ref{sec:no_relocation}).

%%%%%%%%%%%%%%%%%%%%%%%%%%%%%%%%%%%%%%%%%%%%%%%%%%%%%%%%%%%%%%%%%%%%%%%%%%%%%%%%
%
\vspace{-10pt}
\section{Relocation with stackable vehicles}
\label{sec:relocation_stackable}
\noindent
In car sharing systems with stackable cars like ESPRIT~\cite{esprit},  relocation becomes more efficient, since a single relocator can redistribute several vehicles at a time (up to 7, in ESPRIT). But how much more efficient with respect to regular cars and autonomous cars? %The goal of this paper is to provide an initial answer to this question.

In order to evaluate the efficacy of redistributing stackable vehicles, we first need to define a redistribution strategy.  Before discussing the selected approach, we first introduce some terminology. The balance of vehicles at stations can be positive (more available cars than needed with respect to the foreseen demand) or negative (fewer cars than needed). Stations with positive balance are called \emph{feeders}, because they can provide cars to stations in need of cars (which, in turn, are called \emph{recipients}).  Also, we denote with $v_T$ the maximum train size allowed. Similarly to the related literature~\cite{ijrr12_fluid}, we perform relocation periodically (every $T$ minutes). The proposed algorithm follows three steps: i) identify the expected vehicle balance/unbalance, ii) match feeders and recipients, iii) match relocators and feeder-recipient pairs. These steps are discussed in detail below.

%%%%%%%%%%%%%%%%%%%%%%%%%%%%%%%%%%%%%
\subsection{Identify the expected vehicle balance/unbalance}
\label{sec:identifying_roles}
\noindent
First, we want to understand the excess/deficiencies of vehicles in the network. We denote with $b_i(kT)$ the overall balance of vehicles at station $i$ during time interval $[kT, (k+1)T)$ and we write it as follows:
\begin{multline}\label{eq:balance}
b_i(kT) = v_i(kT) + drop_i^{[kT, (k+1)T)} + \\ - pick_i^{[kT, (k+1)T)} - v_i^{control}(kT),
\end{multline}
where $v_i(kT)$ denotes the number of cars parked at station $i$ at time $kT$, $drop_i^{[kT, (k+1)T)}$ and $pick_i^{[kT, (k+1)T)}$ are, respectively, the expected number of drop-offs and pickups in the next time interval \emph{according to the demand and the currently relocating vehicles}. Quantity $v_i^{control}(kT)$ is a control knob that denotes how many vehicles should be present at station~$i$ no matter what (e.g., it can be set to 1 only if the rest of the right-hand side of Equation~\ref{eq:balance} is positive, in order to be conservative in the choice of feeders). Intuitively, the excess/deficiencies of vehicles are computed balancing the expected in/out flows at each station and taking into account the current situation. If $b_i(kT)>0$, station $i$ is expected to have an abundance of vehicles in $[kT, (k+1)T)$, hence we can use these vehicles for relocation. Vice versa, if $b_i(kT)<0$, station~$i$ is expected to have a deficiency of vehicles in $[kT, (k+1)T)$. Thus, we identify as \emph{feeders} all the stations for which $b_i(kT)>0$ (and we denote their set as $\mathcal{F}$). Analogously, we define the set of recipients $\mathcal{R} = \{ i  : b_i(kT)<0 \}$.

More complex strategies for demand prediction could be developed, e.g., exploiting machine learning techniques. However, to the purpose of the paper, suffices to use the same approach for all the relocation strategies considered, in order to have a fair comparison.

%%%%%%%%%%%%%%%%%%%%%%%%%%%%%%%%%%%%%
\subsection{Match feeders and recipients}
\label{sec:matching_feeders_recipients}
\noindent
We now want to assign to each recipient one or more feeders in order to address its unbalance. This can be achieved in several ways. The idea behind our approach is that, since relocators are costly, we want use as few of them as possible. We have the two sets of feeders ($\mathcal{F}$) and recipients ($\mathcal{R}$), each ordered in decreasing order of cars in excess/missing\footnote{In the following, we drop the reference to $kT$, as it is implicit that we are referring to the current relocation interval. }. We define $V_{excess} = \sum_{j \in \mathcal{F}} b_j$ and $V_{deficit} = -\sum_{i \in \mathcal{R}} b_i$. Please note that if $V_{excess} \geq V_{deficit}$, then all deficits can be addressed, at least in principle. Vice versa, if $V_{excess} \leq V_{deficit}$, we know that we are not able to fulfil the requests. Then, we proceed according to the pseudo-code in Algorithm~\ref{alg:greedy_feeder_recipient} below, which must be run every $T$. The main idea behind it is to first address the unbalance of the most unbalanced recipients (those with the smaller $b_i$, which is negative for feeders) using the most rich feeders. If the needs of the recipient are not completely satisfied using one feeder, the recipient is put back in the list of recipients with unfulfilled requests, with an updated balance. Similarly, if the excess of cars at a feeder are not exhausted by a single recipient, the feeder is kept in $\mathcal{F}$ but its balance is updated.

\setlength{\textfloatsep}{0pt}
\begin{algorithm}
\small
		\begin{algorithmic}[1] 
			\Require $\mathcal{R}$ set of recipients, $\mathcal{F}$ set of feeders, $v_T$ max train size
			\Ensure $\mathcal{P}$ set of feeder-recipient pairs that balance the network
			\Statex
			\State $\mathcal{P} = \{ \}$
			\While {($\mathcal{R}\neq\emptyset \land \mathcal{F}\neq\emptyset $) }
				\State find recipient $r_i \in \mathcal{R}$ such that $b_{r_i}$ is the smallest in $\mathcal{R}$
				\State find feeder $f_j \in \mathcal{F}$ such that $b_{f_i}$ is the greatest in $\mathcal{F}$ \Comment{\parbox[t]{0.3\linewidth}{Ties are broken picking the one with the smallest $T_{f_{j},r_{i}}$}}
				\If {$b_{r_i} < b_{f_j}$} \Comment{Recipient can be fully satisfied, feeder still with excess}
					\State $v = \min\{b_{r_j}, v_T\}$ \Comment{Relocate at most $v_T$ vehicles}
					\State $\mathcal{P} = \mathcal{P} \cup \{ \left< f_j,  r_i, v \right>\}$ \Comment{Add the pair to the result}
					\State $b_{f_j} = b_{f_j} - v$ \Comment{Update the excess at the feeder}
					\State sort $\mathcal{F}$ \Comment{Sort the feeders list}
					\State $b_{r_j} = b_{r_j} - v$ \Comment{Update the deficit at the recipient}
					\If { $b_{r_j} == 0$ }
						\State $\mathcal{R} = \mathcal{R} - \{ r_i \}$ \Comment{...remove the recipient}
					\Else
						\State sort $\mathcal{R}$
					\EndIf
				\ElsIf {$b_{r_i} == b_{f_j}$} \Comment{Perfect match}
					\State $v = \min\{b_{r_j}, v_T\}$
					\State $\mathcal{P} = \mathcal{P} \cup \{ \left< f_j,  r_i, v \right>\}$ \Comment{Add the pair to the result}
					\State $b_{f_j} = b_{f_j} - v$ \Comment{Update the excess at the feeder}
					\State $b_{r_j} = b_{r_j} - v$ \Comment{Update the deficit at the recipient}
					\If { $b_{r_j} == 0$ }
						\State $\mathcal{R} = \mathcal{R} - \{ r_i \}$ \Comment{...remove the recipient}
						\State $\mathcal{F} = \mathcal{F} - \{ f_j \}$ \Comment{...remove the feeder}
					\Else
						\State sort $\mathcal{F}$ and $\mathcal{R}$ 
					\EndIf
				\ElsIf {$b_{r_i} > b_{f_j}$} \Comment{Recipient can't be fully satisfied}
					\State $v= \min\{b_{f_j}, v_T\}$
					\State $\mathcal{P} = \mathcal{P} \cup \{ \left< f_j,  r_i, v \right>\}$ \Comment{Add the pair to the result}
					\State $b_{f_j} = b_{f_j} - v$ \Comment{Update the excess at the feeder}
					\State $b_{r_j} = b_{r_j} - v$ \Comment{Update the deficit at the recipient}
					\State sort $\mathcal{R}$
					\If { $b_{f_j} == 0$ }
						\State $\mathcal{F} = \mathcal{F} - \{ f_j \}$ \Comment{Remove the feeder}
					\Else
						\State sort $\mathcal{F}$ 
					\EndIf
				\EndIf
			\EndWhile
			\Statex
			\State \Return $\mathcal{P}$
		\end{algorithmic}
\caption{Feeder-recipient matching algorithm}\label{alg:greedy_feeder_recipient}
\end{algorithm}

%%%%%%%%%%%%%%%%%%%%%%%%%%%%%%%%%%%%%
%\vspace{-10pt}
\subsection{Match relocators and feeder-recipient pairs}
\label{sec:matching_relocators_and_feeders_recipients_pairs}
\noindent
The output of the previous step provides us with a list of feeder-recipient pairs and the number of cars to be relocated between them. Now we have to assign one relocator to each of these pairs. In the following, we denote with $\mathcal{P}$ the set of matched feeder-recipient pairs and with $\mathcal{O}$ the set of relocators. Please note that if $| \mathcal{P}| > |\mathcal{O}|$ the relocation demand might not be satisfied (unless some operators finish their task quickly and carry out another relocation within the same $T$). The idea is to assign relocators in order to minimise the overall relocation time (composed of the time for the relocator to reach the feeder from its current position plus the time to go from feeder to recipient). Since relocators may not be enough to satisfy all relocation requests, we prioritise relocation tasks that move the most vehicles. This approach is summarised in the Algorithm~\ref{alg:greedy_relocator_feeder_recipient}. Please note that Algorithm~\ref{alg:greedy_relocator_feeder_recipient} can be run more frequently than every $T$ (possibly, as soon as there is an idle relocator available).

\setlength{\textfloatsep}{0pt}
\begin{algorithm}[h]
\small
		\begin{algorithmic}[1] 
			\Require $\mathcal{P}$ set of feeder-recipient pairs, $\mathcal{O}$ set of relocators
			\Ensure $\mathcal{M}$ set of matched relocator-feeder-recipient triplet
			\Statex
			\State $\mathcal{M} = \{ \}$
			\While {($\mathcal{O}\neq\emptyset \land \mathcal{P}\neq\emptyset $) }
				\State find the pair $p = (f_j, r_i)$ in $\mathcal{P}$ that relocates the most vehicles $v_p$
				\State find the operator $o$ in $\mathcal{O}$ such that $T_{o,f_j} + T_{f_j,r_i}$ is minimum
				\State $\mathcal{M} = \mathcal{M} \cup \{ \left< o, f_j,  r_i, v_p \right>\}$ \Comment{Add the tuple to the result}
				\State $\mathcal{O} = \mathcal{O} - \{ o \}$ \Comment{Remove the operator}
				\State $\mathcal{P} = \mathcal{P} - \{ p \}$ \Comment{Remove the feeder-recipient pair}
			\EndWhile
			\Statex
			\State \Return $\mathcal{M}$
		\end{algorithmic}
\caption{Relocator-Feeder-Recipient matching algorithm}\label{alg:greedy_relocator_feeder_recipient}
\end{algorithm}

%%%%%%%%%%%%%%%%%%%%%%%%%%%%%%%%%%%%%%%%%%%%%%%%%%%%%%%%%%%%%%%%%%%%%%%%%%%%%%%%
%
%\vspace{-10pt}
\section{Relocation with autonomous vehicles}\label{sec:relocation_autonomous} %\vspace{-10pt}
\noindent
As discussed earlier, autonomous vehicles provide the maximum flexibility in terms of relocation, since they can proactively move towards a hot spot when they end up in a cold spot. A simple relocation strategy for autonomous vehicles can be obtained by modifying the algorithms for relocation with stackable vehicles described in Section~\ref{sec:relocation_stackable}. Basically, it is enough to drop step 3 (corresponding to Algorithm~\ref{alg:greedy_relocator_feeder_recipient}), and assume that, for each feeder-recipient pair $p = (f_i, r_j)$, $v_{p}$ vehicles will autonomously move from $f_i$ to $r_j$ at the time of relocation.

%%%%%%%%%%%%%%%%%%%%%%%%%%%%%%%%%%%%%%%%%%%%%%%%%%%%%%%%%%%%%%%%%%%%%%%%%%%%%%%%
%
%\vspace{-10pt}
\section{Evaluation}
\label{sec:evaluation} %\vspace{-10pt}
\noindent
In order to evaluate the performance of relocation based on stackable vehicles, we have developed a custom C++ simulator that models how vehicles move across the stations in the system. The simulator takes as input the car sharing demand discussed in Section~\ref{sec:demand}, so shared cars move between stations according to the pickup and drop-off requests in the demand\footnote{Please note that the goal of the paper is not to reproduce the daily evolution of mobility on the transport network of the study area but to study, in simplified settings, the potentialities of relocation with stackable vehicles. For this reason, the richness of transport simulators like MATSim~\cite{horni2016multi}, is not needed here. We also neglect the effect of congestion in the road network of the study areas. While more detailed analyses are planned as future work, we argue here that congestion will affect all policies approximately in the same way, and hence this simplified analysis still capture the main aspects of relocation performance.}. The travel times between stations are estimated based on a map of the study area. 

\begin{figure}[H]
\centering
\begin{subfigure}[b]{1\linewidth}
\includegraphics[scale=0.35]{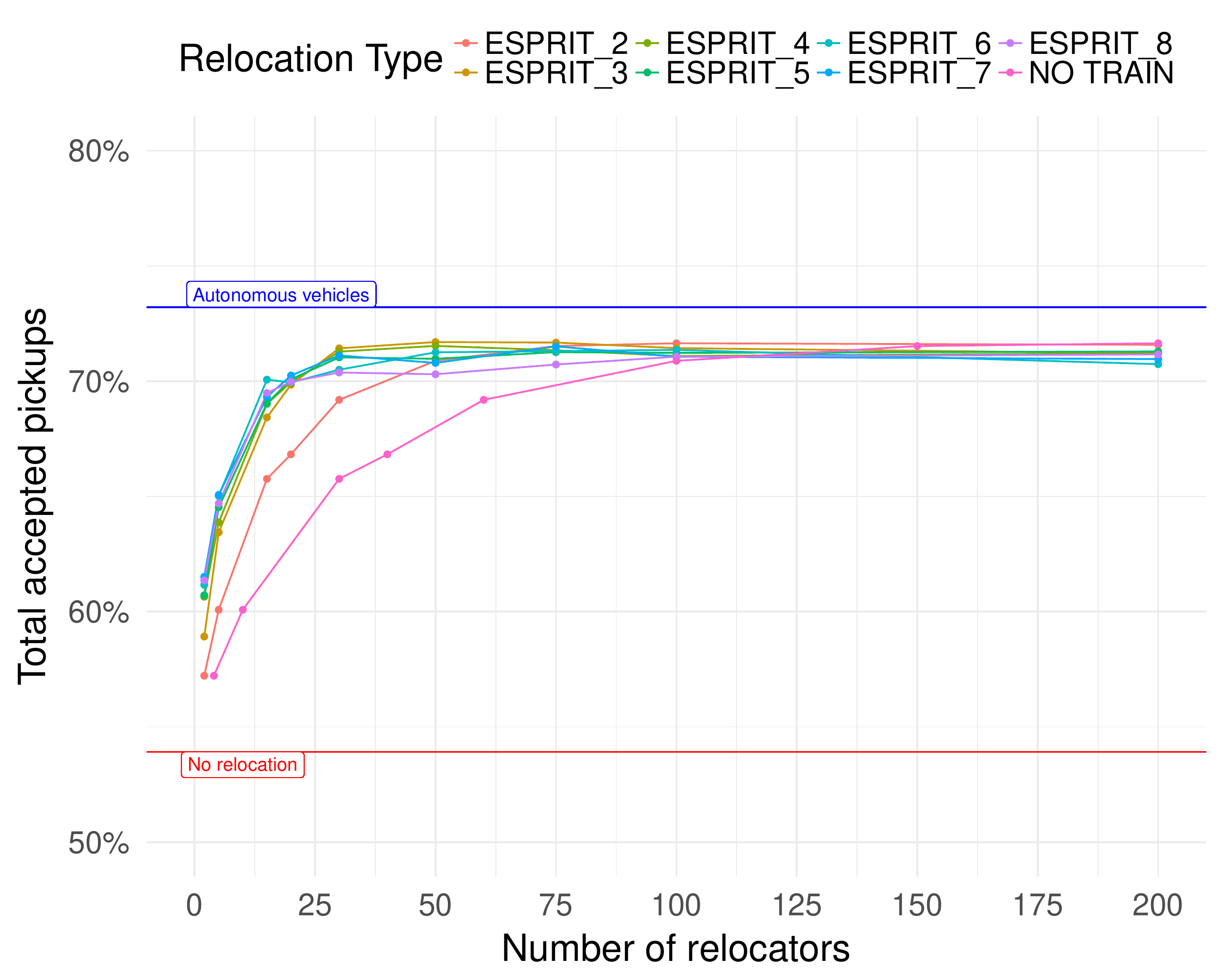}
\caption{Relocation interval = 5 minutes.}
\label{fig:accepted_pickups_300}
\end{subfigure}
\begin{subfigure}[b]{1\linewidth}
\includegraphics[scale=0.35]{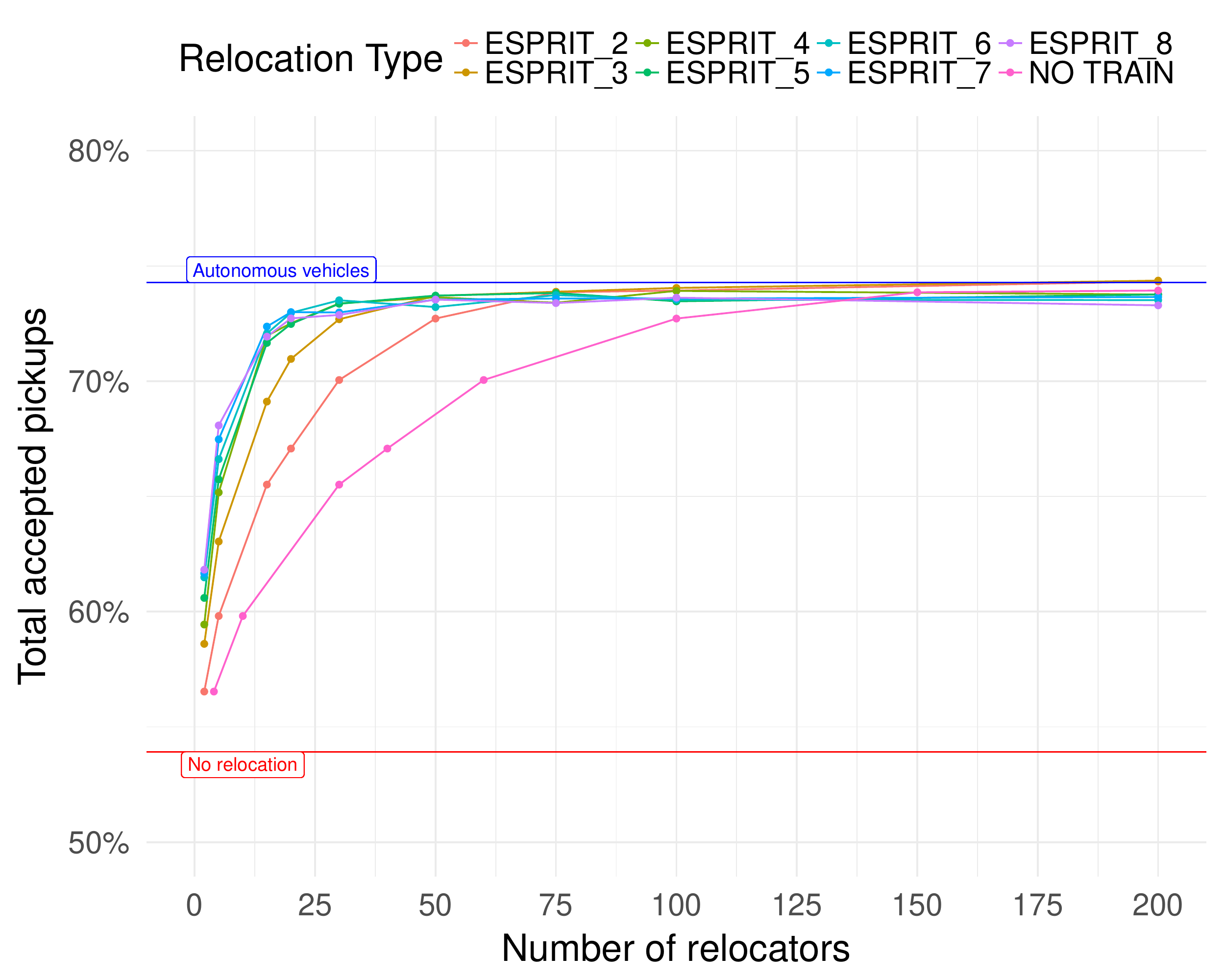}
\caption{Relocation interval = 15 minutes.}
\label{fig:accepted_pickups_900}
\end{subfigure}
\begin{subfigure}[b]{1\linewidth}
\includegraphics[scale=0.35]{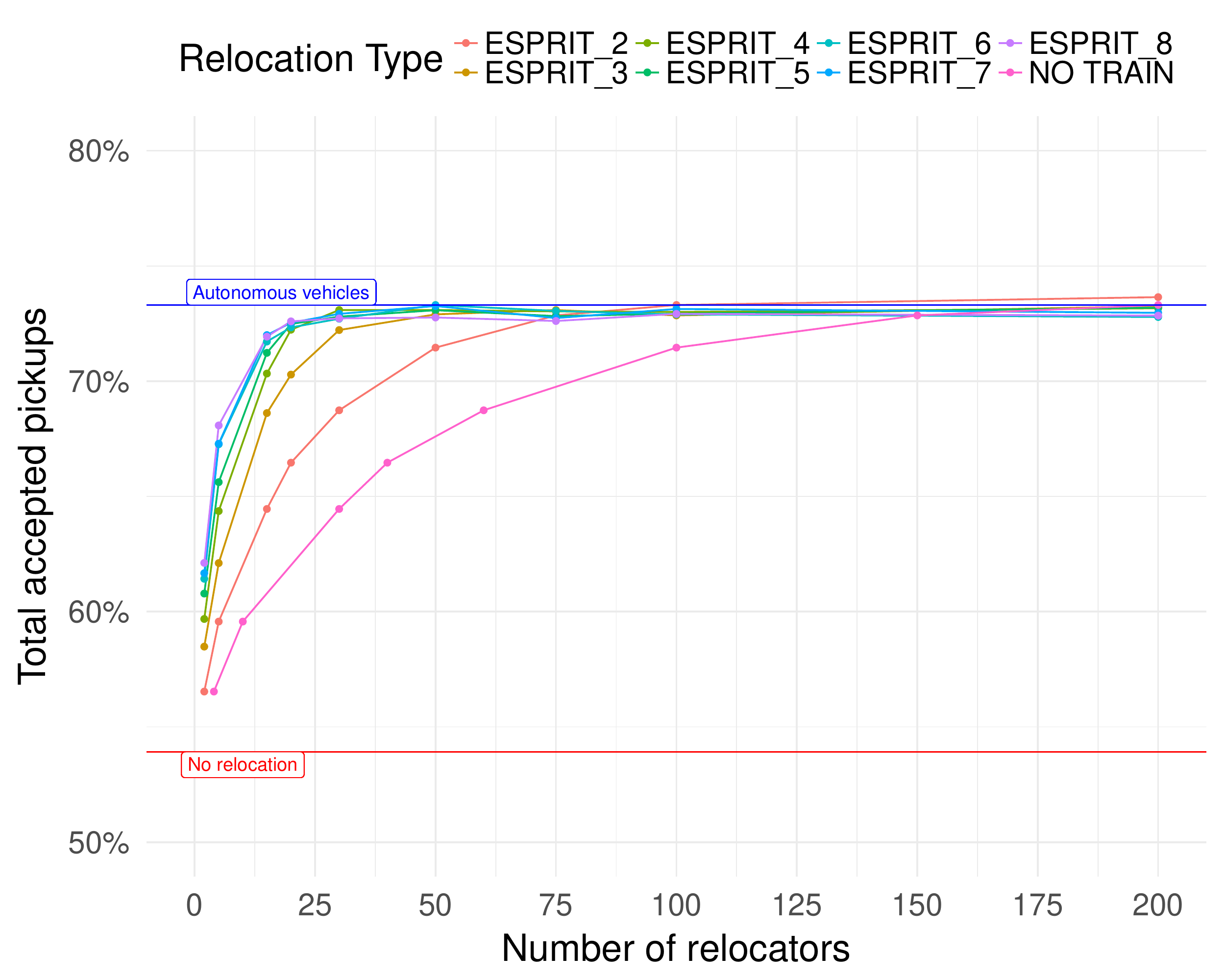}
\caption{Relocation interval = 30 minutes.}
\label{fig:accepted_pickups_1800}
\end{subfigure}
\caption{Accepted pickups}
\label{fig:accepted_pickups}
\end{figure}

On top of this mobility process we deploy the relocation strategies described in Sections~\ref{sec:relocation_stackable}-\ref{sec:relocation_autonomous}. We assume that the fleet size (i.e., the number of cars in the car sharing system) is 350, i.e., a bit more than 292, the number needed according to the theoretical model discussed in Section~\ref{sec:relocation_theo_bound} for perfect rebalancing. Parameter $v_T$ is varied during simulations from 2 to 8.  Please note that a relocator driving a train of 8 vehicles will be able to relocate at most 7, since he/she needs one of the cars in the train to reach the next feeder. Finally, the relocation interval $T$ is varied in $\{5, 15, 30\}$ minutes. 
                                  
The main performance metric considered is the percentage of total accepted requests with respect to the original demand. In Figure~\ref{fig:accepted_pickups} we plot it for different train sizes and relocation intervals, as well as a function of the number of relocators. It is interesting to note that, with a relocation window of 30 minutes and $\sim30$ relocators (Figure~\ref{fig:accepted_pickups_1800}), stackable vehicles provide exactly the same level of service as autonomous cars when the maximum allowed train size is greater than or equal to 5 cars. In all cases, relocating provides better performances than no relocation, but standard relocation needs a lot of relocators before it catches up with autonomous cars. In all cases, even autonomous cars are not able to satisfy 100\% of the demand. There may be several reasons for that. First, the number of vehicles may be too low for the demand. The number of vehicles was chosen so that it was above the threshold obtained from the theoretical model in~\cite{ijrr12_fluid}. However, the model was obtained under simplifying assumptions that may affect the results when cast onto real systems (e.g., real demands have a time-varying behaviour during the day, while the model does not). Second, the demand prediction tool may be too simplistic. In any case, this simple setup is enough to highlight striking differences between the relocation of different types of vehicles.

\begin{figure}[t]
\begin{center}
\hspace{-5pt}\includegraphics[scale=0.33]{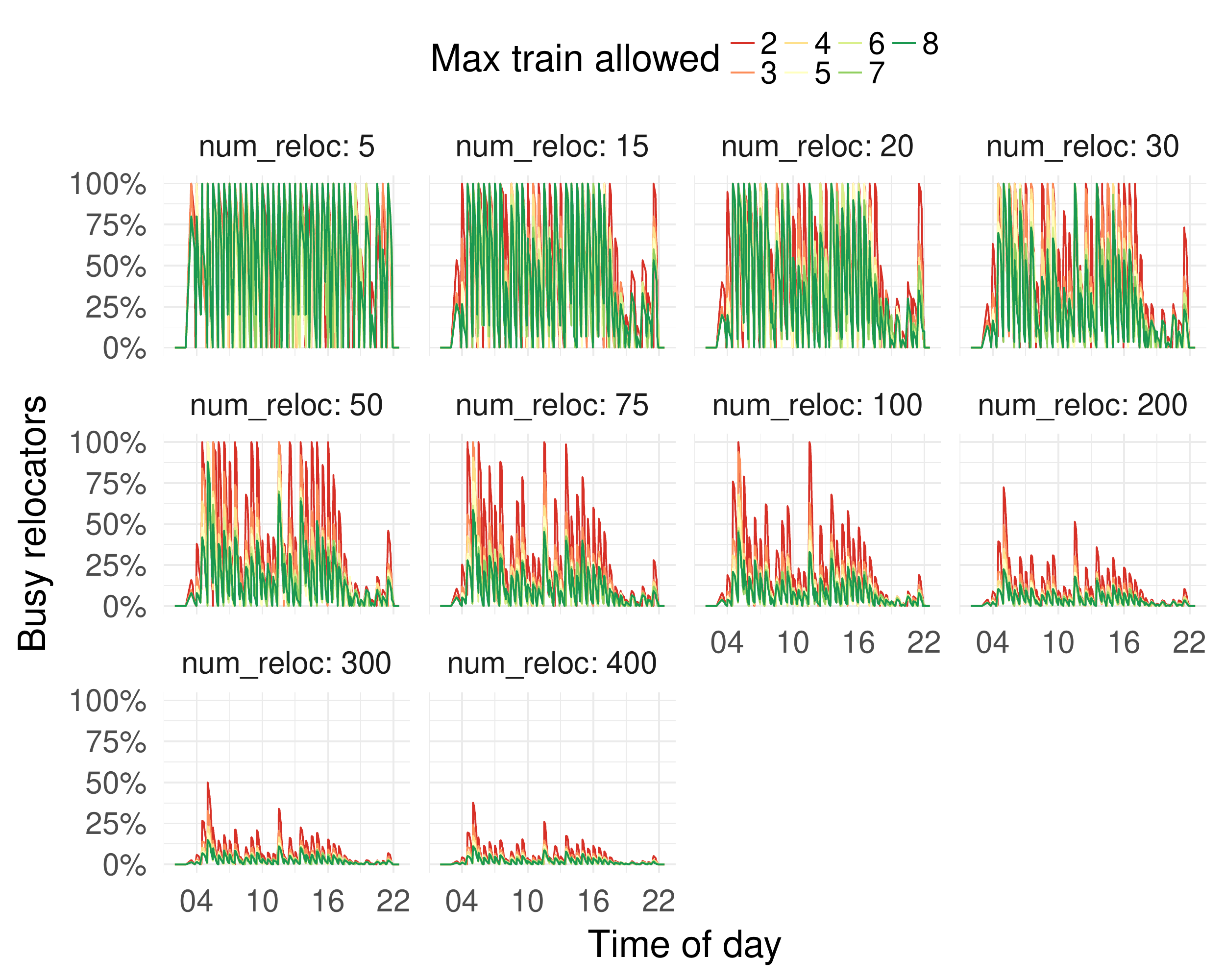}
\caption{Time series of busy relocators (relocation interval = 30 minutes).}
\label{fig:ts_relocators_1800}
\end{center}
\end{figure}

\begin{figure}[t]
\begin{center}
\hspace{-15pt}\includegraphics[scale=0.3]{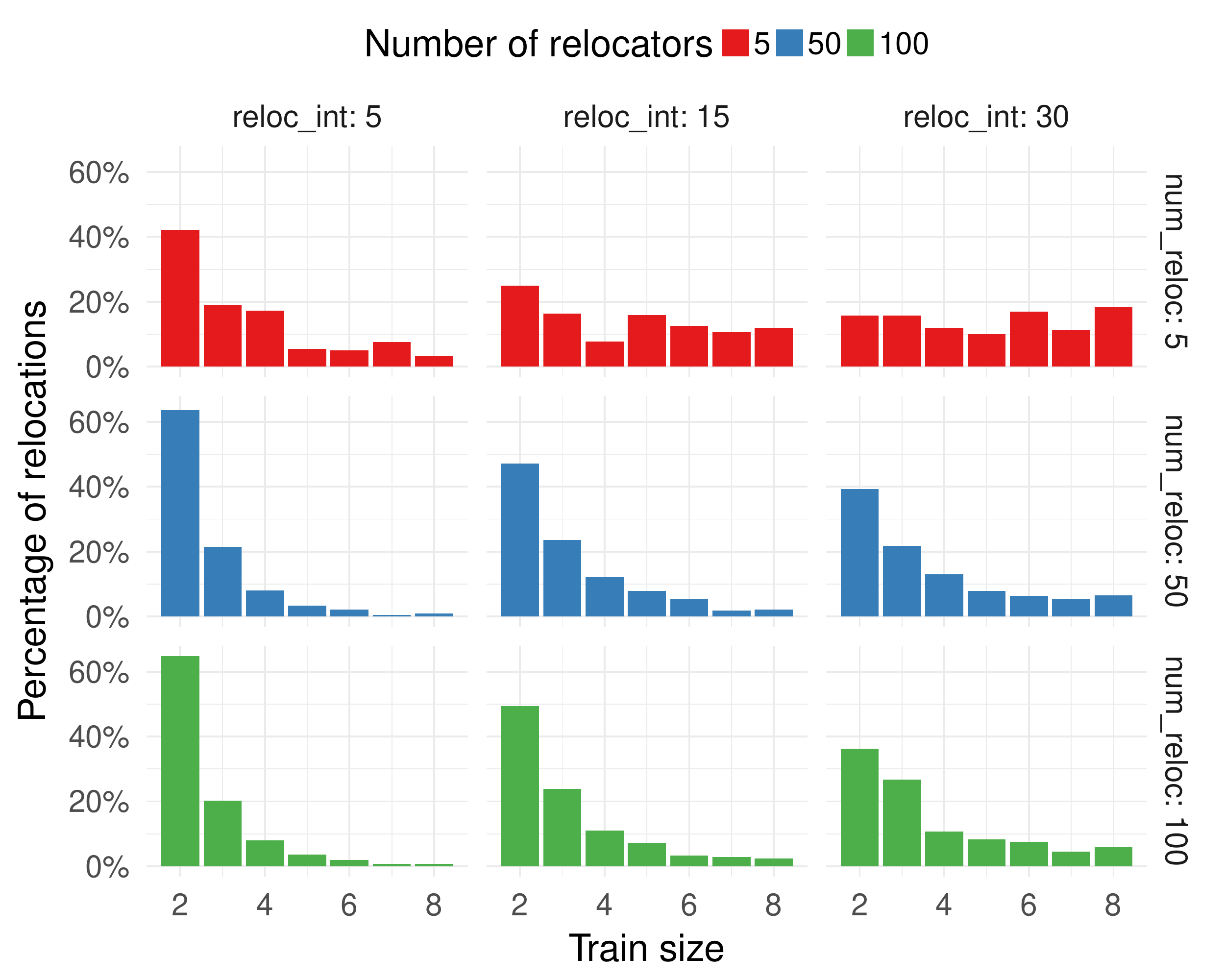}
\caption{The train length.}
\label{fig:train_size}
\end{center}
\end{figure}

A second important metric (especially from the car sharing business model standpoint) is the number of relocators employed to perform relocation, and how long they are busy relocating. We show these quantities in Figure~\ref{fig:ts_relocators_1800} for a relocation interval of 30 minutes. Peaks in relocation activities are clearly visible during the day. If we focus on the case of 30 relocators and maximum train size 8 (the best case scenario discussed above), we observe that only early in the morning the entire relocation workforce is needed. For the rest of the day, only a fraction of the workforce is busy. This is definitely helpful for organising work shifts and deriving the full-time equivalents in the business model.

Finally, we plot the train size distribution as a function of the relocation interval and the number of relocators (Figure~\ref{fig:train_size}). There is a clear trade-off between the two. When relocators are a lot and relocation is performed frequently, relocators mostly use short trains (because the overall relocation burden is low). Vice versa, with just a few relocators and infrequent relocations, long trains are needed to meet all relocation requests.

%%%%%%%%%%%%%%%%%%%%%%%%%%%%%%%%%%%%%%%%%%%%%%%%%%%%%%%%%%%%%%%%%%%%%%%%%%%%%%%%
%
\section{Conclusions}\label{sec:conclusions}
\noindent
In this work we have presented a preliminary analysis, based on a realistic car sharing demand, of the relocation capabilities of stackable vehicles. Our results show that stackable vehicles are able to bridge the relocation gap between autonomous vehicles and standard cars, by exploiting trains of vehicles that can be driven by a single relocator. More specifically, with just 30 relocators, a car sharing system with stackable vehicle achieves the same performance in terms of accepted demand as a car sharing system with self-driving cars. These results show that stackable vehicles can be a cost-effective, efficient solutions while waiting for self-driving cars to become a mainstream reality.
%\vspace{-10pt}
                                  
%%%%%%%%%%%%%%%%%%%%%%%%%%%%%%%%%%%%%%%%%%%%%%%%%%%%%%%%%%%%%%%%%%%%%%%%%%%%%%%%
%
\section*{Acknowledgement}
\noindent
We would like to thank Helen Porter and Peter Davidson for letting us use in this paper the car sharing demand that they have produced for the ESPRIT study area in Lyon.

%This work was partially funded by the ESPRIT project. This project has received funding from the \emph{European Union's Horizon 2020 research and innovation programme} under grant agreement No 653395. This work was also partially funded by the REPLICATE project. This project has received funding from the \emph{European Union's Horizon 2020 research and innovation programme} under grant agreement No 691735.

%
%
%%%%%%%%%%%%%%%%%%%%%%%%%%%%%%%%%%%%%%%%%%%%%%%%%%%%%%%%%%%%%%%%%%%%%%%%%%%%%%%%
%
%\section*{APPENDIX}
%\noindent
%
%
%
%
%%%%%%%%%%%%%%%%%%%%%%%%%%%%%%%%%%%%%%%%%%%%%%%%%%%%%%%%%%%%%%%%%%%%%%%%%%%%%%%%
%
%\section*{ACKNOWLEDGMENT}
%\noindent
%This work was partially funded by the ESPRIT project. This project has received funding from the \emph{European UnionÕs Horizon 2020 research and innovation programme} under grant agreement No 653395.

%%%%%%%%%%%%%%%%%%%%%%%%%%%%%%%%%%%%%%%%%%%%%%%%%%%%%%%%%%%%%%%%%%%%%%%%%%%%%%%%

%\vspace{-0.3cm}
\bibliographystyle{IEEEtran}

\bibliography{mod17.bib}

\end{document}